# Search for overlapped communities by parallel genetic algorithms

**Vincenza CARCHIOLO**
**Dipartimento di Ingegneria Informatica e delle Telecomunicazioni,**
**Universita' degli Studi di Catania**
**Catania, I95125 ITALY**

**Michele MALGERI**
**Dipartimento di Ingegneria Informatica e delle Telecomunicazioni,**
**Universita' degli Studi di Catania**
**Catania, I95125 ITALY**

**Alessandro LONGHEU**
**Dipartimento di Ingegneria Informatica e delle Telecomunicazioni,**
**Universita' degli Studi di Catania**
**Catania, I95125 ITALY**

**Giuseppe MANGIONI**
**Dipartimento di Ingegneria Informatica e delle Telecomunicazioni,**
**Universita' degli Studi di Catania**
**Catania, I95125 ITALY**

***Abstract*— In the last decade the broad scope of complex networks has led to a rapid progress. In this area a particular interest has the study of community structures. The analysis of this type of structure requires the formalization of the intuitive concept of community and the definition of indices of goodness for the obtained results. A lot of algorithms has been presented to reach this goal. In particular, an interesting problem is the search of overlapped communities and it is field seems very interesting a solution based on the use of genetic algorithms. The approach discusses in this paper is based on a parallel implementation of a genetic algorithm and shows the performance benefits of this solution.**

***Keywords-component; formatting; style; styling; insert (key words)***

## I. INTRODUCTION

The network concept has become increasingly pervasive in modern society and it was realized that many realities are shaped in the form of networks. In addition to those that intuitively are modeled as networks, such as social networks, internet, www, computer networks, the network concept is perfectly suited to model reality in many different areas from Biomedical, Life Science, Medicine to Business and Economics [1][2].

Network modeling real systems are often characterized by a large amount of nodes and links between them. For this reason the discipline born of that impulse is known as "complex network".

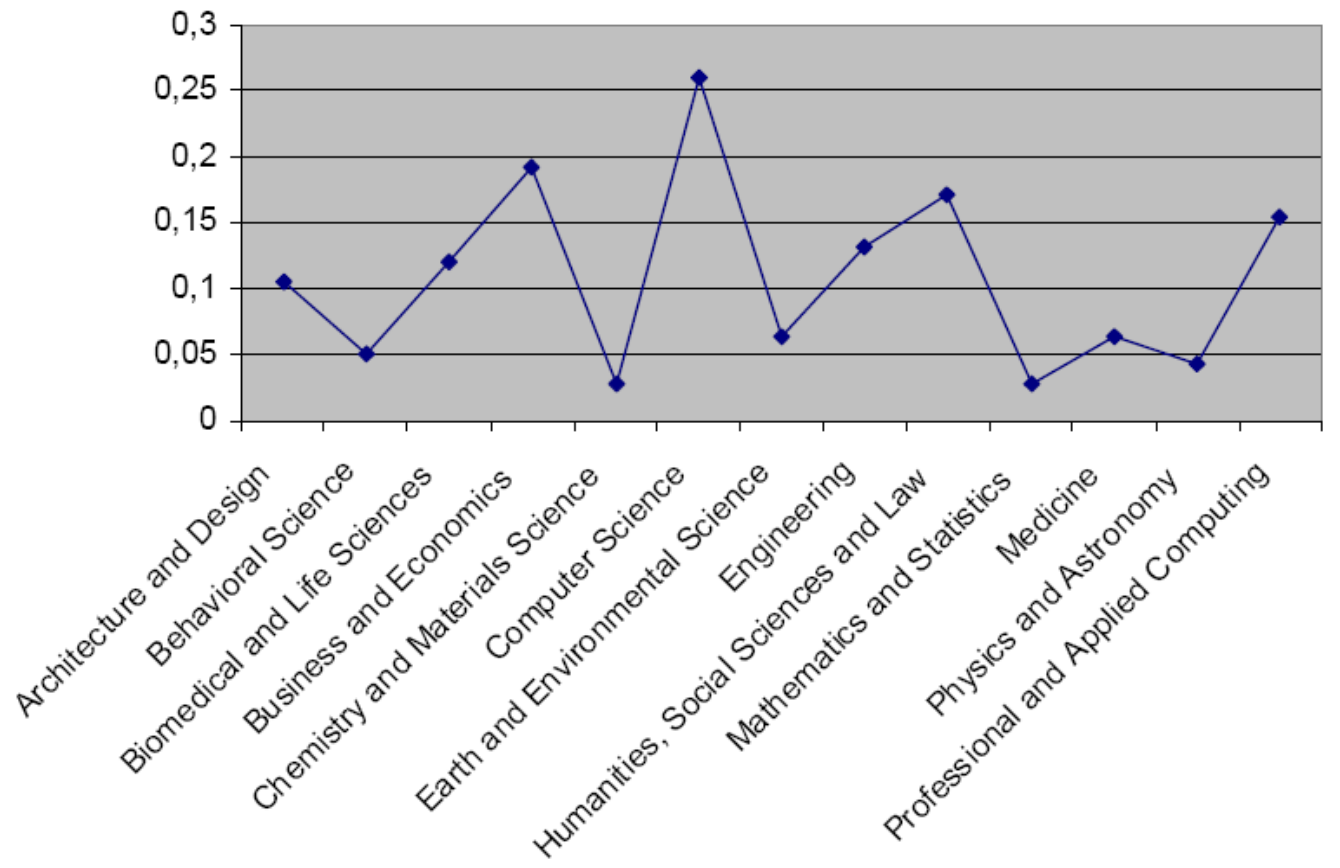

Figure 1. *Percentage of publications in the area of complex network theory*

Therefore, the community of researchers who worked in the field of complex network is gradually enlarged. This has caused a great boost and a broadening of the areas in which studies on networks have been applied.

The search for the sentence "complex networks" in an archive of scientific publications unequivocally shows the major impact of networks in real life. In fact, we will find a huge amount of citations regarding complex networks in several field of interests. As an example, Figure 1 shows the result of search of the string “complex networks” on an online database of scientific publications. This database covers different fields and the figure shows the percentage of publications concerning with complex networks.

Since a complex network can be modeled as a graph, the “complex networks theory” can be seen as derived from the oldest “graph theory” and much of the knowledge from graph theory have been spilled on the study of complex networks. On the other hand, the need to investigate about specific topological and dynamic characteristics arose thus requiring to analyze networks with new tools not present in the classical graph theory. A major propulsion these studies was provided by the pioneering works of Watts and Strogatz [3] and Barabasi and Albert [4].

To define and characterize the properties of these complex networks various measures have been introduced. They defined several topological and dynamic related properties as degree of connectivity, average path length, clustering, etc.

Another structural property that recently has attracted the attention of many scientists is the presence of communities. A **community** is a set of network's nodes that has a greater connectivity among its components respect to rest of the network. Figure 2 shows an example of a network in which it is possible to identify a clear communities structure.

Communities are very important since they are often associated with functional units of a system, like groups of individuals who interact in a society, web pages on the same arguments, and so on. The identification of communities could be considered as a unit a bit rough of the network, however, it provides information on the roles of individual nodes this showing that a complex network is not simply a huge amount of anonymous nodes. For example, a node located between two communities can act as a mediator between them. On the contrary, a node in more central position within a communities provides control and stability to it.

A lot of studies have shown that communities have different characteristics with respect to the entire network. Hence, focus only on the entire network, ignoring this structure, causes a loss of important information about some network characteristics.

The discovering of communities received a considerable attention in recent years and in this field there is still a continuous evolution. This is demonstrated by the numerous algorithms for community structure detection that have been recently proposed (as detailed in the next sections).

The work proposed in this paper aims at presenting a parallel genetic algorithm to discover a specific class of communities named overlapped communities. We will show how community detection can greatly benefit of this parallel implementation achieving high level of performance.

The paper is organized as in the following. Section 2 introduces the concept of community structure and the definition of modularity given by Newman [5]. Section 3 deals with the description of overlapping communities. Section 4 presents the parallel genetic algorithm used in this paper to discovery overlapped communities. Section 5 discusses about the results of the experiments with the proposed algorithm on a set of test case networks.

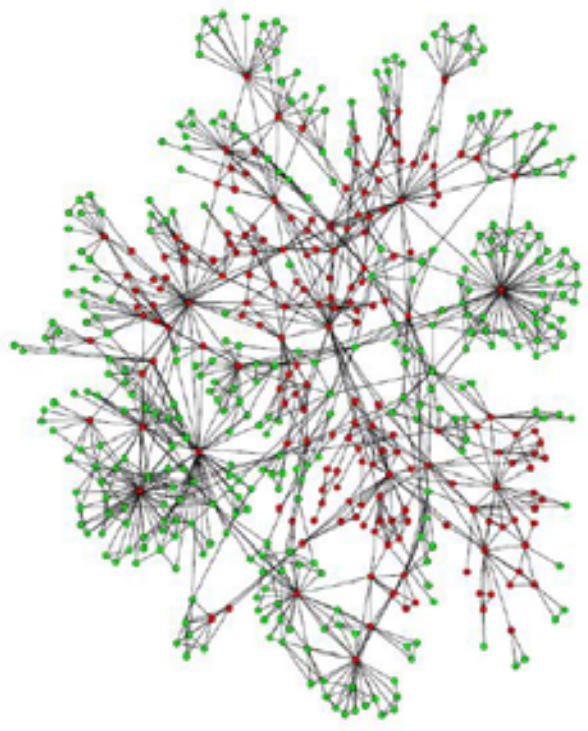

Figure 2. *Community Structure*

## II. Community Structure

The definition of community is influenced by the kind of network studied, in fact links can be weighted or less. Moreover, we can suppose the presence of hierarchies of communities, that is, the communities may, in turn, divided into communities. Furthermore, another aspect in the definition of community concerns the possibility of considering that a nodes can belong simultaneously to different communities or not. In the first case, we referred as **overlapping communities.** In this paper we will take care of this case.

While there is no definition of community that is commonly accepted, there are many works in this direction. Here, following the approach proposed by Fortunato [7], we classified the community definitions into three major categories:

- Local definition
- Global definitions
- Definitions based on vertex similarity

In literature it is presents a lot of criteria to identify communities based on a local view [8] [9] [10]. The main point of these approaches is to focus on a subgraph rather than the entire graph. For example in [9] a sort of subgraph named clique is introduced. It corresponds to a very strong community whose members are connected to each other. More precisely, a clique is defined as a subset of a graph, containing more than two nodes, where all nodes are interconnected through links in both directions; in a clique, the shortest path between all nodes is equal to 1. Several clique-like structures have been defined with different peculiarities in terms of diameter or strangeness of links inside the sub-structure, as for example n-clique [8] [9], n-clan and n-club [10], k-plex [11], k-core [12] and so on.

Global view based community definitions focus on the structural characteristic of the entire network. One common approach is based on the idea that a graph has a community structure if its structure is “different” from that of a given graph used as *null model*. Normally the null model is a random graph, i.e. a graph where links between nodes are placed at random and as a consequence it doesn't display any particular community structure. One of the most popular null model is that proposed by Newman [5], consisting in i) a graph having the same number of nodes of the original graph and ii) each node keeps its original degree and iii) nodes are linked at random. Therefore, a community is defined as a set of nodes within which the number of links is greater than the number of expected links among the same nodes in the null mode. This definition is also used by Newman to define the *modularity* function, a measure that is used as global criterion to define a community and also as a metric function to measure the quality of the partition. The modularity function is upper limited by 1, and the best partitioning is the one with value of modularity nearest to 1. In the case of a "bad" partition , the modularity can also take negative values.

The third category of community definitions is based on the calculation of a similarity function applied to the vertexes. The idea is that nodes that are "similar" belong to the same community. In literature there exist several approaches to community definition based on vertexes similarity and they differ on the choice of the function of similarity.

## III. Modularity and overlapping communities

As discussed in the previous section, the modularity function is a widely accepted measure of the quality of a community partition. The success of modularity definition is mainly due to its simplicity and elegance even if an important resolution limit has been pointed out in [15]. However, given that the original definition of modularity does not cover all kinds of networks, several works have extended the original formulation in order to cope both with weighted [16] and directed graphs [17].

Recently, the modularity function has been further extended to detect overlapping communities [6].

Overlapping communities are very common in real networks, since nodes usually belong to several groups at the same time. For instance in social networks a person usually belongs to several communities since he can have more interests in his life, i.e. he loves to play soccer, to participate in a forum on Internet, etc.. For these reasons, in the formulation of modularity for overlapping communities proposed in [6], it is supposed that a graph node belongs to a community *k* with a strength $\alpha_{i,k}$. Moreover, the authors define a similar factor for each graph link by properly combining belonging factors of link's starting and ending nodes. In particular, given a link $l(i,j)$ from node *i* to node *j*, $\beta_{l,k}= F(\alpha_{i,k},\alpha_{j,k})$ expresses the strength with which link *l* belongs to community *k*. Note that this formulation does not specify which function *F* should be used. Following the approach of [6], the two-dimensional logistic function will be used.

The modularity function $Q_{ov}$ for overlapping communities is then defined as:

$$Q_{ov} = \frac{1}{m}\sum_{c\in C}\sum_{i,j\in V}\left[\beta_{l(i,j),c}A_{ij} - \frac{\beta^{out}_{l(i,j),c}k_i^{out}\beta^{in}_{l(i,j),c}k_j^{in}}{m}\right]$$

where $A_{ij}$ is the adjacency matrix of the network, $k_i^{out}$ and $k_j^{in}$ are respectively the out-degree of node *i* and the in-degree of node *j*. In this definition, the terms $\beta_{l(i,j),k}^{out}$ and $\beta_{l(i,j),k}^{in}$ represent the expected belonging factor of any possible link *l*, respectively, starting from and pointing to a node into community *k*.

## IV. Using genetic algorithms to discovering overlapping communities

While $Q_{ov}$ is essentially a quality function that measures the goodness of a given network partition and, it can also be exploited to discover communities. In fact, since higher values of $Q_{ov}$ correspond to better partitions, it is possible to find the best division in overlapping communities directly optimizing the $Q_{ov}$ function. In [6] $Q_{ov}$ optimization is reformulated as a genetic problem and the best partition is obtained using a genetic algorithm (GA).

GAs have been extensively used in the optimization field given their ability to find a solution of a problem especially in the presence of a very large solution space. GAs are essentially based on the evolution of a set of individuals (called population), where each of them represents a possible solution of the optimization problem. The solution of a given problem is obtained through several simulation steps, usually called epochs. At each epoch, individuals are ordered with respect to increasing values of a fitness function which expresses how much each individual is "close" to the optimum. Then, the better individuals by (i.e. those having the highest fitness value) are included in the next generation, while new individuals are created i) combining the best individuals by using a crossover operator and ii) performing random mutations. These two operations mimic the behaviors of life species, so it comes the term "genetic". For a detailed description of GAs please refer to [18].

GAs have been used in the field of communities detection in [19] to optimize the Newman's modularity. Moreover, GAs have been successfully used in [6] to find the overlapping communities partition that optimizes the $Q_{ov}$ function used as the GA fitness function, since better partitions of the network correspond to higher values of $Q_{ov}$.

As stated in [6], the most critical computation in the GA they propose is the fitness evaluation. This operation has a computational complexity of $O(|C|*n^2)$ in the worst case, where $|C|$ is the number of overlapping communities and *n* is the number of nodes of the network. Such a level of complexity prevents the use of this method especially for large networks.

In the present work, we propose the use of a parallel genetic algorithm to discover overlapping communities. Our proposal aims at reducing the execution time thus permitting to uncover community structures on larger networks. The parallel genetic algorithm we used is based on the single population global model implemented by a *master/slave* algorithm. This means that a process called *master* holds all the population individuals and executes all steps of the genetic algorithm except the fitness evaluations. These last operations are performed by *slave* processes. Communication among *master* process and *slaves* is implemented by message passing, using the MPI library [20]. Since in our problem the most time consuming operation is the fitness evaluation, we greatly benefit from parallel implementation achieving a high level of performance.

Moreover, in order to evaluate the speedup gain provided by the parallel implementation, we compared it with a sequential implementation of the genetic algorithm. It was our care to implement a sequential genetic algorithm that shares as many functions as possible with the parallel implementation, thus permitting to estimate the speedup gain reducing as much as possible implementation specific delays.

## V. SIMULATION RESULTS

In this section we show the results obtained using a parallel GA to discover overlapping communities on directed networks. The library for genetic algorithms PGAPACK [21] was used fot the implementation of this algorithm. We carried out two versions of this algorithm: a sequential and a parallel ones. The latter implementation has required the use of the MPI Message Passing libraries [20].

Experiments were performed on a multiprocessor machine with the following characteristics:

- 4 CPU: Intel(R) Xeon(R) E5345 a 2.33 GHz with 4096 KB cache
- RAM = 1048724 kB
- Linux operating system

Several simulations have been performed on this machine to assess the value of the real speedup related to the number of processors.

In the realization of the two genetic algorithm implementations we used the following parameters:

- population size = 60
- population to replace = 10% of the initial population
- tournament selection method

In order to measure the speedup gain of the parallel implementation with respects to the sequential one, we define the following speedup figure:

$$S = \frac{T_{par}}{T_{seq}}$$

where $T_{seq}$ is the processing time of the sequential algorithm and $T_{par}$ is the processing time of parallel algorithm. In the ideal case S tends to the number of CPUs.

Speedup tests have been carried out with a number of processors ranging from 1 to 4. The cases study used in our experiments are the Zachary Karate Club network [22] and the Dolphins Social network[23]. Zachary Karate Club represents the friendships between members of a karate club and it is modeled as a network with 34 nodes and 156 links (Figure 3).

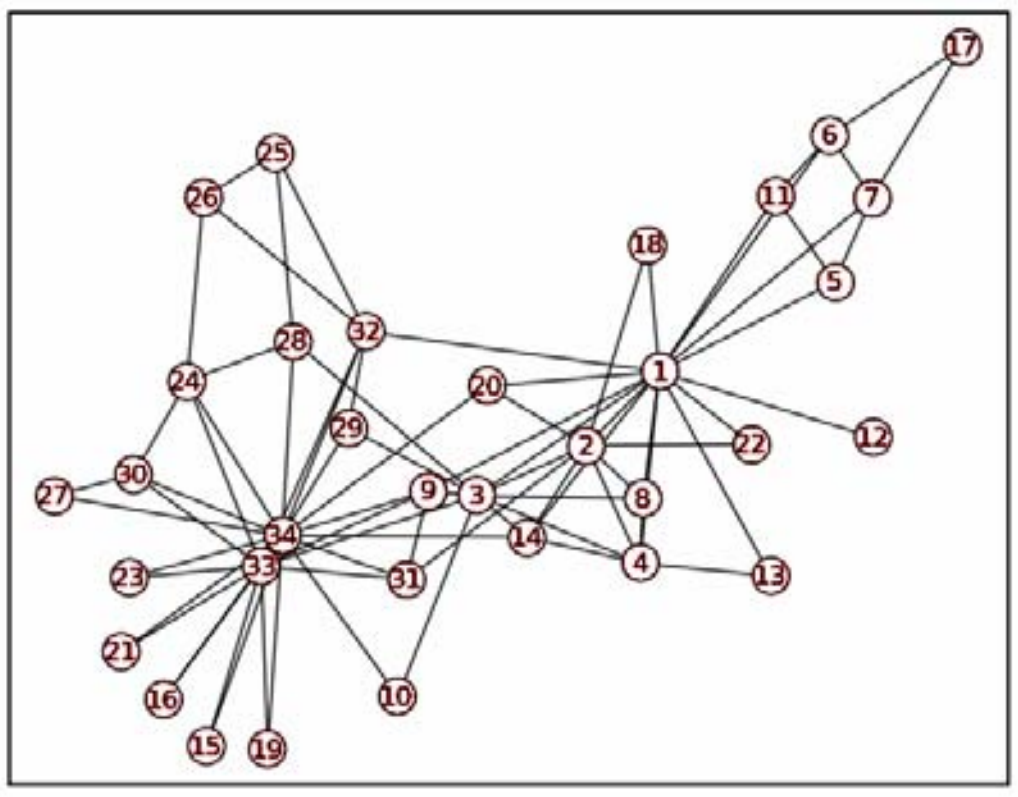

Figure 3. Zachary *Karate Club Network e*

This network has been extensively used as a benchmark for many community discovering algorithms, since on the base of studies made from sociologists it is known that this network is composed by two main communities. The parallel genetic algorithm proposed in this work is able to correctly find the two Zachary network communities, as shown in figure 4. In such a figure, red nodes (i.e. nodes 3 and 10) are found to be overlapped between the two main communities. In particular, $Q_{ov}$ is maximum when node *3* and *10* belong to the green community with a factor respectively of 0.8 and 0.62 (and conversely belong to the yellow community with a factor respectively of *0.2* and *0.38).*

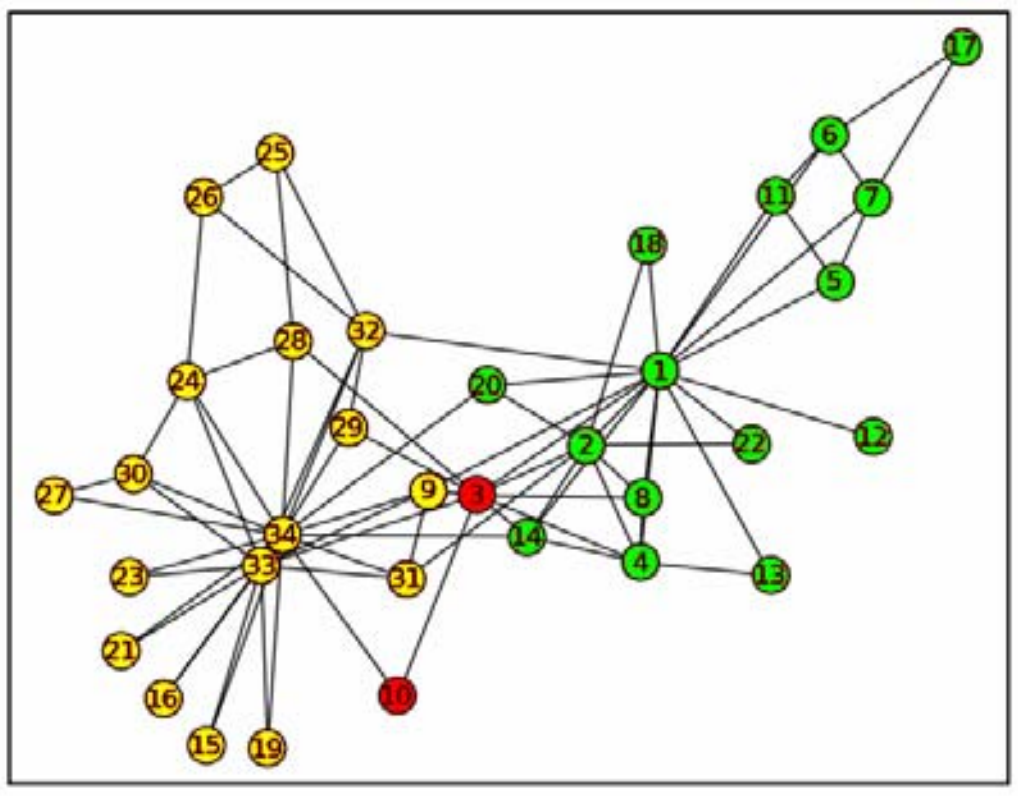

Figure 4. *Best partitioning of Zachary Karate Club*

The Dolphins Social Network rapresents is a social network that represents the constant companion of dolphins. This network contains 62 nodes and 318 links (figure 5). This network has also been deeply studied by biologists and it is known that it presents four communities, as shown in figure 6. In particular, red and blue nodes represents females dolphins, while green and white are for male dolphins. Even in this case, the proposed algorithm correctly identifies the four communities.

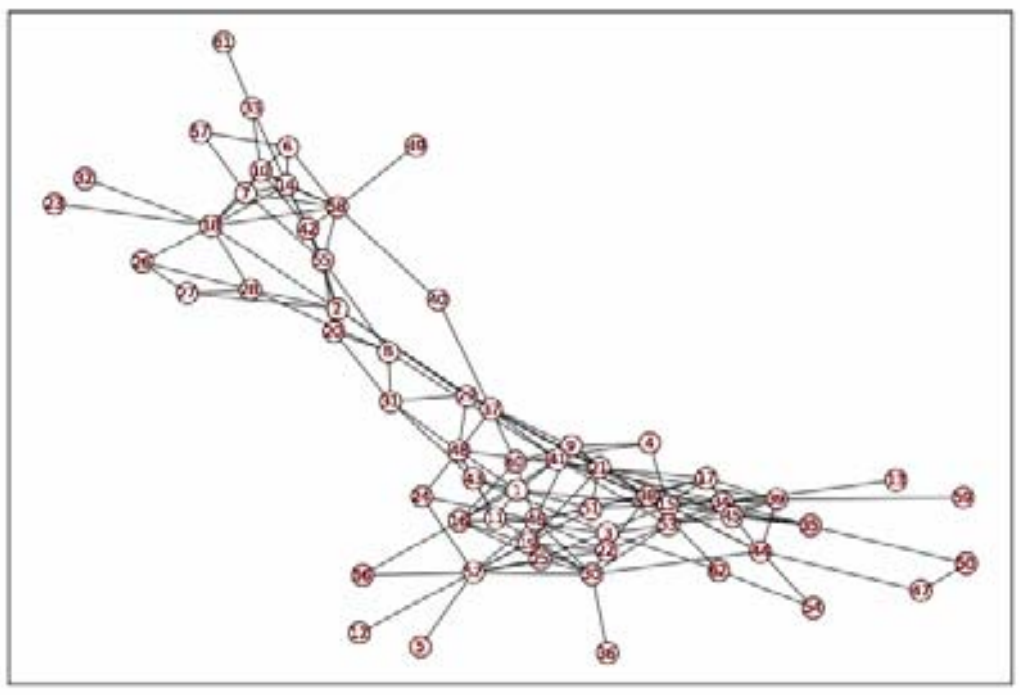

Figure 5. - *Dolphin social netwok*

Several tests have been carried out for each combination of processors number in order to analyze the performance gain of the proposed parallel GA implementation.

Speedup S for the Zachary network case is shown if Figure 7. Let's note that the speedup obtained with 4 processors is almost identical to that obtained with 3 processors. This can be explained by the implementation of PGAPACK. Indeed, for combinations of up to 3 processors, the evaluation of individuals is are carried out by both slave and master. When processors are greater or equal 4 the evaluation of individuals is only delegated to slaves. In both cases, therefore, the processors involved in the fitness evaluation operations are 3.

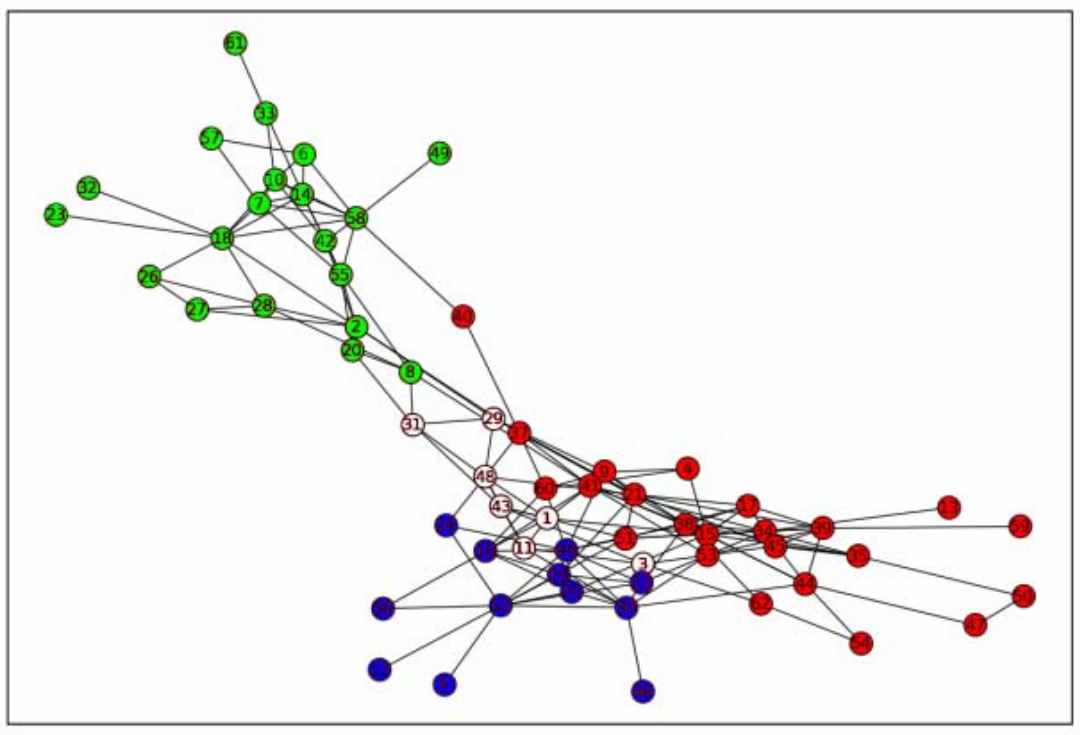

Figure 6. *Communities in Dolphin Social Network*

Figure 8 shows the speedup for Dolphins Social Network.

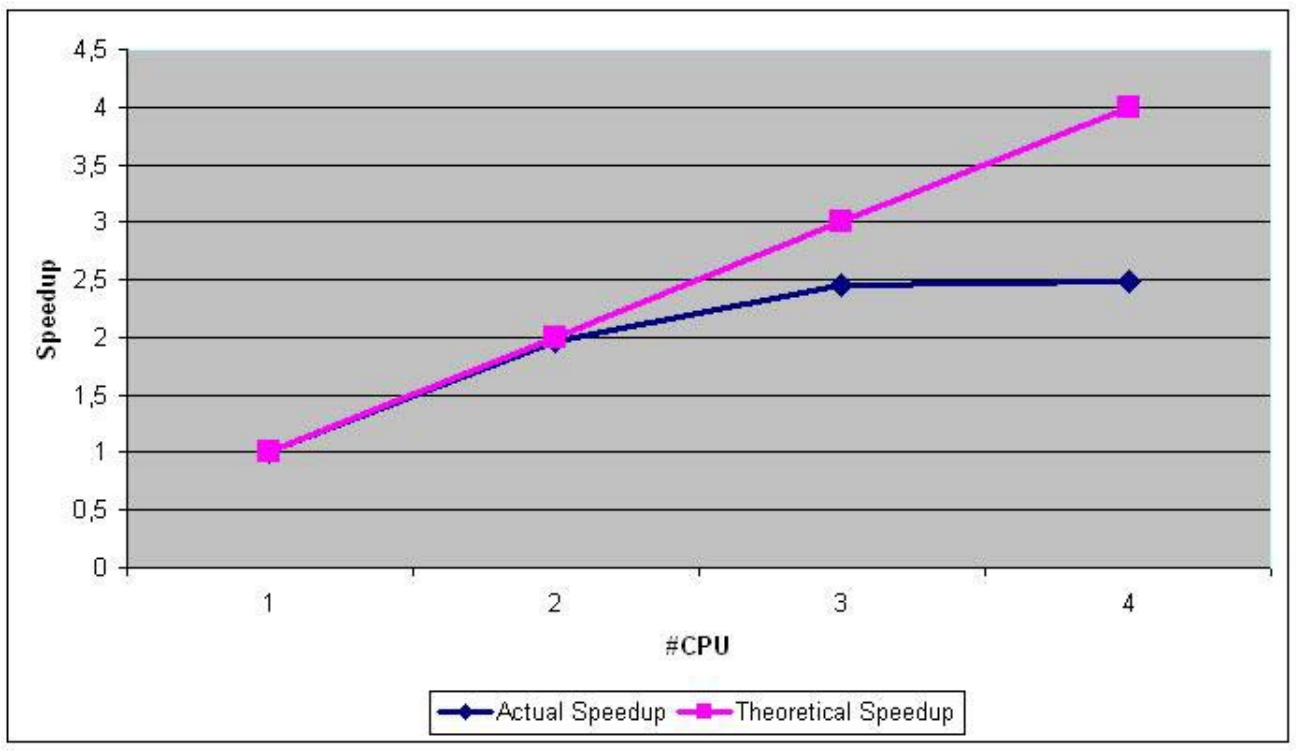

Figure 7. *Speedup for the Zachary Karate Club*

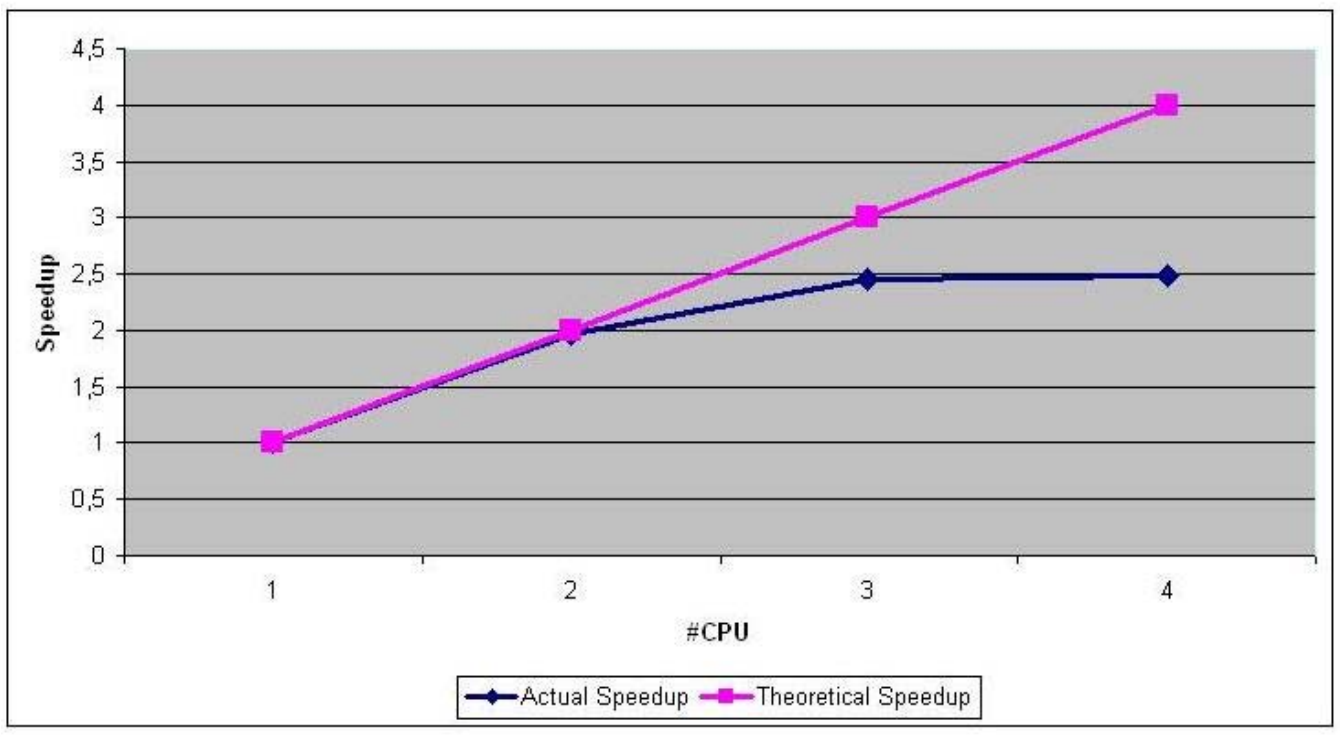

Figure 8. *Speedup for Dolphins Social Network*

Also in this case, performances obtained using 3 or 4 processors are almost the same. In general, we can conclude that, as it was expected, the speedup gain is proportional to the number of processors dedicated to the execution of slave processes, since these are the most time consuming tasks. Looking at figures 7 and 8, we also note that when the number of processors greater than 2, *S* is approximately proportional to *N*0.8*, where *N* is the number of processors.

## VI. Conclusions

The problem of communities discovery in a network has widely discussed in the paper with particular attention at the case of overlapped communities. Among several algorithms to communities discovering we have chosen to implement the one presented in [6]. In order to improve performance we chose to provided a parallel implementation of that algorithm.

In order to measure the performance of that solution a measure of speedup respect to the sequential solution was defined and the implementation has been tested on Zachary Karate Club network and Dolphins Social Network.

The results seem very promising and we intend perform further study with more complex networks using architectures with a larger number of processors.

## AUTHORS PROFILE

**Vincenza Carchiolo** is currently full of Computer Science in Department of Informatics and Telecommunications at University of Catania. Her research interests include information retrieval, query languages, distributed system, and formal language. She received a degree with Honors in Electrical Engineering from University of Catania, Italy in 1983.

**Alessandro Longheu** received his MS in Computer Engineering in 1997 from the University of Catania and then his PhD in 2001 from the University of Palermo. He currently teaches Programming Languages and he was also a Professor of Computer Networks at the Faculty of Engineering of Catania. His research interests include e-learning, workflows, information retrieval and integration in the semantic web, complex networks and trust and semantic web.

**Giuseppe Mangioni** is assistant professor in Department of Informatics and Telecommunications at University of Catania. He received the degree in Computer Engineering (1995) and the Ph.D. degree (2000) at the University of Catania. Currently he is professor of Computer Networks at the Faculty of Engineering of Catania. His research interests include peer-to-peer systems, trust and reputation systems, self-organizing and self-adaptive systems and complex networks.

**Michele Malgeri** is associate professor in Department of Informatics and Telecommunications at University of Catania. His research interests include distributed system, information retrieval, query languages and formal language. He received a degree with Honors in Electrical Engineering from University of Catania, Italy in 1983.

…